\documentclass[a4paper]{jpconf}
\usepackage{graphicx}
\usepackage{floatrow}

\input epsf

\def\plotonesmall#1{\centering \leavevmode
\epsfxsize=0.7\columnwidth \epsfbox{#1}}

\begin{document}
\title{Ukaliq: Seeing Long-Term with Small, Precise Arctic Telescopes}

\author{Eric Steinbring, Brian Leckie \& Rick Murowinski}

\address{National Research Council Canada, Herzberg Astronomy and Astrophysics, 5071 West Saanich Road, Victoria, British Columbia, Canada V9E 2E7 }

\ead{Eric.Steinbring@nrc-cnrc.gc.ca}

\begin{abstract}
Time-domain astrophysics benefits from extreme-latitude sites, which can combine intrinsically extended nighttime with good sky conditions. One such location is the Polar Environment Atmospheric Research Laboratory (PEARL), at $80^{\circ}$ North latitude, on the northwestern edge of Ellesmere Island, Canada. Experience gained deploying seeing monitors there has been incorporated into an automated system called ``Ukaliq" after the common arctic hare, which is also very well suited to its local environment. Even with modest aperture, high photometric reliability may be achieved using simple adaptive optics together with observing strategies that best fit the unique set of advantages available at PEARL: excellent image quality maintained during many clear, calm, dark periods of 100 hours or more. A potential multi-year search for gravitational microlensing of quasars with Ukaliq helps illustrate this niche in the era of large wide-field survey facilities.
\end{abstract}

\section{Introduction}

High-precision optical/near-infrared photometry uninterrupted for many hours to days, and reliably repeated over months to years is a special observing cadence for which polar sites can provide a distinct advantage. This paper presents a concept that could allow a cluster of identical optical telescopes of up to 0.5 m aperture, for both site testing and science, to be deployed in the High Arctic. The combined aperture would be equivalent to a 1.2 m telescope. We call the unit systems ``Ukaliq", the Inuktitut word for the abundant arctic hare. Included are adaptations specialized to the atmospheric statistics, particularly cloud cover (ice crystals) and boundary-layer seeing, which depend on local wind conditions. To show the good fit, an instructive example of dedicating one unit to monitoring quasar microlensing events is provided.

\section{Site Environment}

Near the Eureka weatherstation at $80^{\circ}$ North latitude on Ellesmere Island, Canada, is the Polar Environment Atmospheric Research Laboratory (PEARL). It was designed for atmospheric studies with optical instruments and has a purpose-built observing platform on its flat roof: 7 m wide by 18 m across, with an external stairwell. Over the last 5 years this has supported astronomical observations, with new ones planned [1], possible in part thanks to the Canadian Network for Detection of Atmospheric Change (CANDAC) which maintains other instruments there. The observatory is situated on a 600 m-high ridge, accessed by 4x4 trucks along a 15 km-long road from the sea-level base facility.  That is operated by the federal weather service of Environment Canada (EC) and serviced by an all-weather airstrip and a yearly resupply ship.

All-sky camera images obtained from PEARL over several winters show no clouds overhead 48\% of the time, 68\% was clear (extinction in $V<0.5$ mag) with a mean duration near 100 hours, and 84\% was deemed usable ($V<2$ mag) [2]. Air temperatures are typically $-30$ C with light winds and often subject to suspended ice crystals, sometimes referred to as ``diamond dust".  These periods are interrupted by brief, but sometimes intense, storms. Winds can gust to over $30~{\rm m}~{\rm s}^{-1}$, usually from the north or south. Free seeing measured during several roughly week-long winter campaigns with Multi-Aperture Scintillation Sensor (MASS) had a mode near 0.23 arcsec and a median of 0.50 arcsec.  It is possibly poorer when winds are completely calm. The total seeing measured from 8 m elevation with MASS combined with Differential Image Motion Monitors (MASS/DIMM) on the roof had a median of 0.76 arcsec. Seeing is best near the median ground windspeed of $4~{\rm m}~{\rm s}^{-1}$, degrading with higher windspeeds [3].

\section{Instrument Heritage and Design}

The Ukaliq approach is based on experience gained deploying MASS and DIMMs at PEARL using Meade 25 cm and 35 cm LX200 ACF optical tube assemblies (OTAs). A larger, 50 cm version is also available from Meade. These are relatively inexpensive, lightweight OTAs easily handled by a crew of one to three people.  All are f/8 Schmidt-Cassegrain (or Ritchey-Chretien) OTAs with a field corrector lens at the aperture into which is mounted a central secondary (with approximately $20$\% obscuration fraction). A sealed OTA has the nice property of keeping ice crystals (snow) from entering the tube. A downside is that this sometimes ``sticky" and optically thick contaminant can then build up along with a thin layer of frost.

In MASS and DIMM operation, the OTA was almost always pointed at Polaris, that is, within ${10}^{\circ}$ of zenith from PEARL, which may exacerbate the problem in two ways.  First, the aperture corrector plate is pointed almost straight up.  Even a long dew shield or insulated baffle would not be useful in completely avoiding radiation of that surface directly to the cold sky. And second, precipitation of diamond dust is then falling perpendicular to the surface. Both situations might be improved somewhat for telescope positions at lower elevations, e.g. a grazing incidence of falling crystals, but neither would be entirely avoided.

Although not as critical for DIMM, measurements with MASS require strict control over the clarity of the aperture window, and image quality (IQ). These parameters were well monitored, in part because Polaris is a multiple system. In the simultaneous DIMM data, the fainter component Polaris B showed stable IQ over 10 arcsec scales with estimates of Strehl ratio (difference between a stellar image from that of a perfect diffraction pattern) over 60\% with the 25 cm OTA. If similar for the 0.5 m OTA, the point-spread function (PSF) under best seeing could be maintained at 0.8 arcsec full-width at half maximum (FWHM).

\subsection{Aperture Clearing and Enclosure}

One method for clearing the aperture might be to continuously heat the corrector lens by sending a low current through either loops of conductive wire or a transparent metallic surface coating, similar to the method employed by the three 50 cm Antarctic Survey Telescopes (AST3) at Dome A, Antarctica. This is undesirable for DIMM operation, as even very small temperature differences within the OTA are well known to disturb the seeing near a telescope. For most of the PEARL seeing measurements, methanol and wipes were used periodically to remove frost and ice crystals, a highly manual process not easily automated.  

Experimentation was made with radiative heating and blowers, ``heat guns", employed at intervals of about an hour or two. Melting was effective, as was using slight heat to encourage sublimation. Results were satisfactory for DIMM operations, that is, only a short period was needed to clear the aperture ($\sim 2$ min) with no detectable, lingering effect on seeing. The aperture window and OTA are low mass and equilibrate quickly in the open air. Best results came from using an electrical radiative coil element, basically an infrared heatlamp pointing out of an enclosure, over which the OTA could be pointed when needed.

\begin{figure}
\plotonesmall{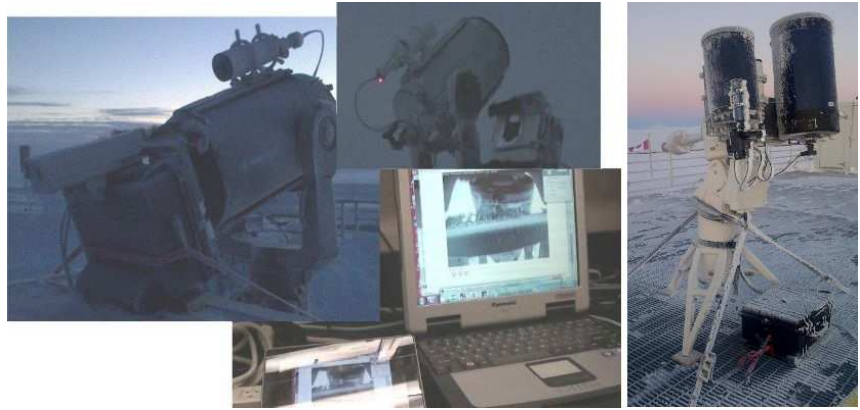}
\caption{The Ukaliq parking box (left and top-middle) includes an electrical radiator to sublimate or melt frost and ice crystals, a soft brush which can be used to remove larger buildup mechanically, a blower, and a webcamera for inspecting the aperture - all controlled remotely (bottom). Discussion of operation with an improved equatorial drive system (right) follows.}
\label{figure1}
\end{figure}

This ``parking box" (pictured in Figure~\ref{figure1}) also doubled as the enclosure for the telescope aperture during high winds. A choice of orientation (into the prevailing wind) presents a minimal profile. With suitable anchorage to the rooftop platform this has provided sufficient protection for the OTA and instruments even during the worst winter storm conditions, and so it was found that Ukaliq does not require a dome. That avoids complexity and cost, and by presenting no extraneous obstacles also preserves the best seeing, which is known to occur during light winds (near the median windspeed) over the PEARL roof.

The reason that sublimation and melting of crystals and frost by radiative heating may work better at PEARL than by other methods is possibly that deposition of ice crystals there is mainly associated with periods of windblown dry snow, not from precipitation when calm.  Frost tends to build up slowly under clear skies, which are known to be associated with calm winds. It also grows very uniformly over the aperture, which allows correction by employing good observing technique with careful relative stellar photometry. Thus, much of the time (when it is clear) aperture clearing is not needed. Otherwise, it is likely to be cloudy anyway. This is also found to be correlated with seeing  poorer than 1 arcsec, that is, winds over $8~{\rm m}~{\rm s}^{-1}$.

\subsection{Autonomous Weatherstation}

Experience with deploying robotic instruments in the Arctic has taught that knowledge of changing weather and a quick response to it are key to preventing damage. The primary issue is wind, not the cold; it is always extremely dry, with relative humidity near saturation. The PEARL facility autonomous weatherstation (AWS) employs a sonic anemometer and is installed at 10 m elevation on a mast in the centre of the roof. It reports averaged conditions via the PEARL network on 10-minute intervals via a webpage. That system is reliable, but for redundancy a separate, dedicated AWS is also integral to Ukaliq. This uses a Young hydrophobic-plastic propeller anemometer and windvane with a Campbell Scientific CR10X datalogger; a common unit used in the High Arctic and previously employed with the ``Inuksuit" devices.

\subsection{Mount and Observing Platform}

Replacing the original Meade LX200 mounts, the Ukaliq drive is an Astro-Physics GTO 3600 German equatorial mount. This was the same drive used by the Antarctic Search for Transiting Exoplanets (ASTEP) 40 cm telescope at Dome C and later, the Dunlap Institute Arctic Telescope (DIAT) 50 cm telescope intended for PEARL. Like those, the Ukaliq drive is lubricated with low-temperature grease. Software from the manufacturer can account for periodic drive errors and drift, easily yielding blind pointing accuracies within a few arcseconds. Several parking positions can be pre-programmed, one of which is set to point the telescope at the parking box. For Ukaliq, an improvement to the ``stock" drive is to replace its mechanical limit switches with those of the magnetic proximity type, avoiding a failure mode caused by packed ice crystals.

A transversely mounted plate, the largest available from Astro-Physics, is used to allow attaching various OTAs on the drive. This always includes a course guiding telescope: an Orion 8 cm ShortTube with a StarShooter USB-based CCD tracker. The mount can accommodate up to two more small OTAs at the same time; both the 25 cm and 35 cm have been installed together, as shown in Figure~\ref{figure1}. These OTAs are each attached to pre-balanced dovetail plates, which mate to matching clamps, so all are interchangeable on the mount. Alternatively, by rotating the plate, a single 50 cm OTA could be mounted with the guider in tandem.

The pier is a squat portable tripod, custom built to fit the mount for the latitude of PEARL. The observing platform is subject to some vibration, and the most stable footing is achieved by placing the tripod over one of three I-beam members running under the floor. The pier is anchored to the members with high-tension cargo straps. The total weight of OTA plus mount and pier is approximately 250 kg and there is potentially floorspace for 5 more of these mounts on the roof. An alternative location (allowing only one unit) is to install on a 6-m tall tower further to the north along the ridge, already prepared for but not installed.

\subsection{Adaptive Optics, Camera and Filter Wheel}

The DIMM data show that mount vibration is minimized when winds are calm, under $4~{\rm m}~{\rm s}^{-1}$, and crew are not on the roof. Under these circumstances, image quality is dominated by seeing, which is still typically subarcsecond. To maintain image quality for windspeeds up to $8~{\rm m}~{\rm s}^{-1}$, when rooftop vibration dominates, each camera would be equipped with its own tip-tilt guiding system. A device could be readily built to meet the need: under 10 arcsec of throw at 10 Hz.
 
A commercial system already meeting similar requirements is the Santa Barbara Instruments Group (SBIG) AO-L combined with an ST-10 camera.  We have not tested that, but have used this camera successfully in the ``Ukpik" system. A scientific instrument might have standard $V$, $R$, and $I$ filters and a 1k X 1k detector. This gives a field of view of about 40 arcsec on a side at 0.4 arcsec/pixel, or larger via a field lens. The built-in guiding detector in the focal plane would then be aligned with a bright ($R<8$ mag) star within 16 arcmin by rotating the field, allowing fast guiding entirely internal to the camera, operated with only a start/stop command.

There is usually no need to protect cameras and electronics under the weather conditions at PEARL; waste heat has been sufficient to keep the SBIG and DIMM  electronics operating, along with the MASS.  Continuously cycling a filter wheel on a duty cycle of minutes can help ensure it remains ``unstuck."  Motors such as those in fans can fail, and are not needed, and so are to be shut off. On occasion, if instruments have been cold-soaked, an application of some heat to achieve normal operating temperature again was enough to restart them. The parking box can be positioned in such a way as to allow it to be used for this procedure as well.

\subsection{Communications and Electronics}

Communication between cameras, mount, and AWS is via RS232 or USB-to-ethernet converters, although direct communication with USB extender cables has also been successful. The bandwidth needed is modest: DIMM observations at 5 Hz rates generated 1 Tbyte of data in 10 days, although an imaging system with a similar number of pixels taking exposures once per minute would be less than that over a full winter.

Telescope electronics (other than instruments) are co-located at the base of the pier, protected inside a separate insulated ``warmbox," somewhat larger than that shown in Figure~\ref{figure1}. For Ukaliq, this box is also the shipping crate for the drive, which is small enough when disassembled to fit inside. When deployed, it houses a supervisory Linux-based single-board computer and USB to ethernet converters.  Power is supplied from outlets on the roof.  Inside the warmbox are power converters to supply 5 V, 12 V and 18 V power to the camera electronics, communications and the telescope drive.  The data-storage unit is kept inside the warm building and communication to the outside world is via the PEARL network and its satellite link.

\subsection{Logistics, Power Consumption and Operations}

Each Ukaliq unit is a clone: OTA, instrument, mount, drive, pier, electronics, warmbox and parking box. Based on previous experience, it would take two to three days per unit to commission on sky. Once deployed in the early winter, the crew would leave the site, and the facility would be autonomous until sunup. Because the cost of crewing PEARL is high, it can be more cost effective to produce and ship duplicate systems (particularly by sea) rather than to maintain a single unit which may fail. During winter, there is limited technical support for fixing things. There has been success with repairs that require no technical expertise. For example, a simple swap of an electronics crate without need to open or adjust it could be feasible.

An attractive feature of Ukaliq is that it employs completely independent, swappable OTAs. Each has its own instrument pre-focused and ready to be deployed on the mount. An identical spare guider allows replacement within minutes. An experiment undertaken during MASS/DIMM observations at PEARL was to focus a duplicate 25 cm OTA for a colder temperature than typical, and later exchange it (when that temperature was reached again) without the need to refocus. The 25 cm OTA is relatively light, and can even be handled by one person, although larger versions would require a crew of just two or three to service. 

Even for a full system of 6 units, operations are simple, and power consumption modest. The majority of the time, each unit is continuously viewing a single field, which for the equatorial mount requires only tracking in right ascension. The guiders are sufficient for acquisition of the target and guidestar for the tip-tilt system. Camera exposures would likely be in a uniform sequence, cycling through the common filter set, and sustained whenever on sky. A warning of winds over about $8~{\rm m}~{\rm s}^{-1}$ or deviating from the prevailing northerlies from the AWS autonomously puts the telescopes into a safe mode; utilizing the parking boxes. A remote observer could then make a decision about resuming operations when safe to do so, first inspecting the apertures to ensure they are clear, or utilizing the parking boxes to make it so. Each radiant element can consume up to 100 W. The drive and electronics are under 10 W, so the maximum total power requirement would be 660 W, but typically below 60 W.

\section{Example Survey Science}

Each unit could independently track a single field for either site testing or science. One well-known gravitationally lensed system which illustrates the good fit of Ukaliq is Q0957+561. It is a doubly imaged quasar at $z=1.405$ lensed by a foreground, $z=0.355$ galaxy. Images A and B are relatively bright, $V=16$, and separated by 6 arcsec on the sky; IQ of 1 arcsec PSF FWHM ensures no blending between them. Their time delay is $417\pm 2$ days, that is, a path difference causes any change in the brightness of quasar image B to lag A, which has some intrinsic color dependence inherent to the source [4 and references therein]. It is remarkable that this is not known more accurately after roughly 30 years of monitoring, but that is in part because of uncertainty due to microlensing by substructure in the lensing galaxy. Since one such 0.25 mag event in 1981 to 1986 there have been several (disputed) detections in the last 20 years with amplitudes below 20 millimag and time periods shorter than about 5 days.  A possibility is that variability over various timescales are superimposed. Ukaliq is an ideal tool to search for that due to its high observing efficiency. It can allow reliable, uninterrupted temporal coverage during successive dark periods, in roughly 100 hour blocks with clear skies and good seeing.

\begin{figure}
\floatbox[{\capbeside\thisfloatsetup{capbesideposition={right,top},capbesidewidth=4.5cm}}]{figure}[\FBwidth]
{\caption{A simulated microlensing event in the lensed quasar Q0957+561 on Julian Day 2449706, lasting 150 hours. Based on the actual recorded PEARL sky conditions at the time, Ukaliq could have seen this and at least half of the time-lagged photometric fluctuation of the (non-microlensed) image the following winter. The short peak lasts 19 hours, requiring efficient round-the-clock observations to sample fully.}\label{figure2}}
{\includegraphics[width=9cm]{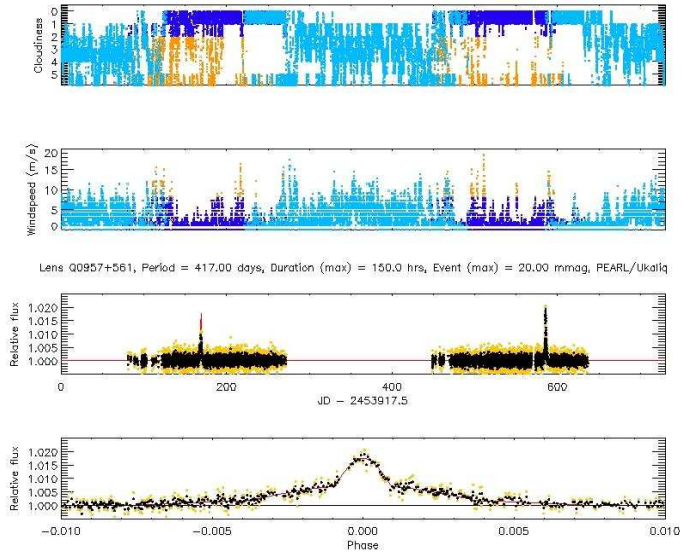}}
\end{figure}

A simple simulation was developed to investigate potential detections. The results are shown in Figure~\ref{figure2}: an artificial signal consistent with a microlensing event for QSO 0957+561 occurring in 2007, the A-B difference lightcurve. It is composed of three co-added symmetric sinusoidal components of 10, 5, and 2 millimags over time periods of 19, 75, and 150 hours.  Real data for sky clarity and windspeed at PEARL were applied, using those for the years 2006-2008. Samples which were too cloudy ($>2.5$) and windy ($>8~{\rm m}~{\rm s}^{-1}$) were excluded (shaded light brown) as have times with civil twilight or brighter (light blue). A realistic photometric error is roughly 1 millimag per combined hourly sample per filter, primarily from Poisson noise. Under these circumstances Ukaliq would detect the peak of the pulse and even be sensitive to the longest timescale fluctuation, because about half of the total duration is always viewed. Shifting the epoch of the event (until it occurred in daylight, not shown) does not change results; the same could be expected in subsequent winters. The full complement of 6 units, equivalent to 1.2 m aperture, could track 5 similar systems as well. Results on one of those for a network of mid-latitude sites might be more like the lightly-shaded (yellow) points, assuming the same average sky conditions, but incurring a penalty of 1 millimag of irreducible systematic error between sites. Even if monitored over 7 years, a much larger aperture at a single mid-latitude site would not help with the shortest-period component, i.e. night/day windowing of the Large Synoptic Survey Telescope (LSST) cannot allow sampling the full lightcurve (if this target was visible).

We thank Mark Halman for expertise in warmbox construction and that of Ivan Wevers during cold testing in the NRC-Herzberg environmental chamber. Helpful discussions with Nicholas Law and Suresh Sivanandam related their knowledge of the mount used with DIAT, as did Jerome Maire through operation of our prototype. Pierre Fogal and Jim Drummond of CANDAC kindly assisted in deploying instruments at PEARL, and we greatly appreciate the support of EC and the hospitality of the weatherstation staff during our Eureka observing runs.

\smallskip

\end{document}